\documentclass{PoS}

\usepackage{subfig}

\title{On the Combined Analysis of Muon Shower Size and Depth of Shower Maximum}

\ShortTitle{Combined Analysis of $N_{\mu}$ and $X_{\rm max}$ for UHECR showers}

\author{\speaker{Jakub V\'icha}, Petr Tr\'avn\'i\v{c}ek\\
        Institute of Physics of the Academy of Sciences of the Czech Republic, \\Na Slovance 2, 182 21 Prague 8\\
        E-mail: \email{vicha@fzu.cz}}

\author{Dalibor Nosek\\
        Faculty of Mathematics and Physics, Charles University in Prague, \\V Hole\v{s}ovi\v{c}k\'{a}ch 2, 180 00 Prague 8}

\abstract{The mass composition of ultra-high energy cosmic rays can be studied from the distributions of the depth of shower maximum and/or the muon shower size. Here, we study the dependence of the mean muon shower size on the depth of shower maximum in detail. Air showers induced by protons and iron nuclei were simulated with two models of hadronic interactions already tuned with LHC data (run I-II). The generated air showers were combined to obtain various types of mass composition of the primary beam. We investigated the shape of the functional dependence of the mean muon shower size on the depth of shower maximum and its dependency on the composition mixture. Fitting this dependence we can derive the primary fractions and the muon rescaling factor with a statistical uncertainty at a level of few percent. The difference between the reconstructed primary fractions is below 20\% when different models are considered. The difference in the muon shower size between the two models was observed to be around 6\%.}

\FullConference{The 34th International Cosmic Ray Conference,\\
		30 July - 6 August, 2015\\
		The Hague, The Netherlands}

\begin{document}

\section{Introduction}
The mass composition of ultra--high energy cosmic rays (UHECR) inducing extensive air showers of energies above $10^{18}$~eV can be measured by fluorescence detectors on average basis. The measurement of the depth of shower maximum ($X_{\rm max}$) is compared with Monte Carlo (MC) predictions in such cases \cite{AugerXmax,TAXmax}. The fluorescence technique provides a precise measurement of $X_{\rm max}$, but a large systematic uncertainty remains in the determination of the mass composition of UHECR. The obstacle comes predominantly from different predictions of hadronic interaction models that are extrapolated from accelerator energies to energies larger by a few orders of magnitude in the center of mass system. The mass composition of UHECR remains uncertain, and even unknown at the highest energies where a steep decrease of the flux is observed \cite{AugerSpectrum,TASpectrum}. The very low statistics of events collected by fluorescence detectors is due to their low duty cycle. 

Assuming a small number of primaries to be present in UHECR, the most probable fractions of these primaries were inferred in \cite{AugerFractions}. The measured distributions of $X_{\rm max}$ were compared with $X_{\rm max}$ distributions by combining MC distributions of the assumed primaries. Large differences in the results were found among the hadronic interaction models. Also, a degeneracy of solutions with similar probability can be expected as, generally, there are many combinations of MC distributions of the individual primaries which describe the measured distributions similarly well.

Whereas the fluorescence technique measures the longitudinal profile of the electromagnetic component of the shower, a measurement of the number of muons on the ground ($N_{\mu}$) can provide an independent way to infer the mass composition of UHECR. Muon detectors have a 100\% duty cycle and, when a good resolution of $N_{\mu}$ is achieved ($\lesssim$10\%), even better separability of the individual primaries can be achieved than in the case of the fluorescence technique. However, there is a lack of $N_{\mu}$ in MC simulations when compared with the measured data \cite{AugerNmuInclined,AugerNmuExcess,AugerNmuExcess2}. The underestimation of muon production is usually characterized in terms of a muon rescaling factor. Moreover, a stronger relationship between $N_{\mu}$ and the shower energy than between $X_{\rm max}$ and the shower energy~\cite{MathewsModel} makes the situation more difficult and an independent measurement of the shower energy is needed for composition studies using $N_{\mu}$. Therefore a simple comparison of the measured distributions of muons with MC predictions would be complicated. 

A combined measurement of UHECR showers with the fluorescence technique ($X_{\rm max}$ and shower energy) and muon detectors ($N_{\mu}$) could be a more sucessfull way to determine the mass composition of UHECR. In the previous studies, the detected muon and electromagnetic signals were utilized to determine the average mass number of a set of air showers, see e.g.~\cite{KascadeLnA}. There are also methods estimating the spread of masses in the UHECR primary beams via correlation of $N_{\mu}$ and $X_{\rm max}$ \cite{CorrelationPaper} or from signals in muon and electromagnetic detectors \cite{SpreadFromSds}. However, it needs to be mentioned that the two currently operating experiments do not yet directly measure the muonic component for showers with zenith angles below 60$^{\circ}$.

Here we present another method to determine the fractions of the assumed primaries in which the rescaling of $N_{\mu}$ ($R_{\mu}$) can be achieved simultaneously with a single fit of a combined measurement of $X_{\rm max}$ and $N_{\mu}$. For this purpose we use MC showers generated with two hadronic interaction models tuned to the LHC data (run I-II).

In the next section, the generated showers that were used in this work are described. The method to determine the fractions of the assumed primaries and $R_{\mu}$ is introduced in Section~\ref{secMethod}. Section~\ref{secResults} contains applications of the method on several examples, testing its performance. The work is summarized in the last section.

\section{Simulated Showers}
For this work we have simulated $\sim10^{5}$ showers with the CONEX 4.37 generator \cite{conex,conex2} for p and Fe primaries with fixed energy $10^{18.5}$~eV and for each of the two hadronic interaction models (QGSJet~II--04 \cite{QGSJETII04} and EPOS--LHC \cite{epos,epos2}). The zenith angles ($\Theta$) of showers were distributed uniformly in $\cos^{2}\Theta$ for $\Theta$ in $<0,~60^{\circ}>$.

Muons with threshold energy 300~MeV at $\sim$1400~m a.s.l. were included to calculate $N_{\mu}$. Electromagnetic particles of energies above 1~MeV formed the longitudinal profile (dependence of the deposited energy on the atmospheric depth), from which $X_{\rm max}$ was fitted with the Gaisser--Hillas function by the program CONEX.

For each shower we used Gaussian smearing of $X_{\rm max}$ and $N_{\mu}$ with a variance equal to $\sigma(X_{\rm max})$ 
and $\sigma(N_{\mu})$, respectively. These smearings imitate the detector resolutions. We adopted a correction for the attenuation of $N_{\mu}$ with zenith angle due to the different amount of atmosphere penetrated by the air shower before it reaches the ground. The correction was made using a polynomial of 3$^{\rm rd}$ order in $\cos^{2}\Theta$ for each model of hadronic interactions. An equally mixed composition of p and Fe was considered for this purpose.

\begin{figure}[h!]
  \centering
\subfloat{\includegraphics[width=0.5\textwidth]{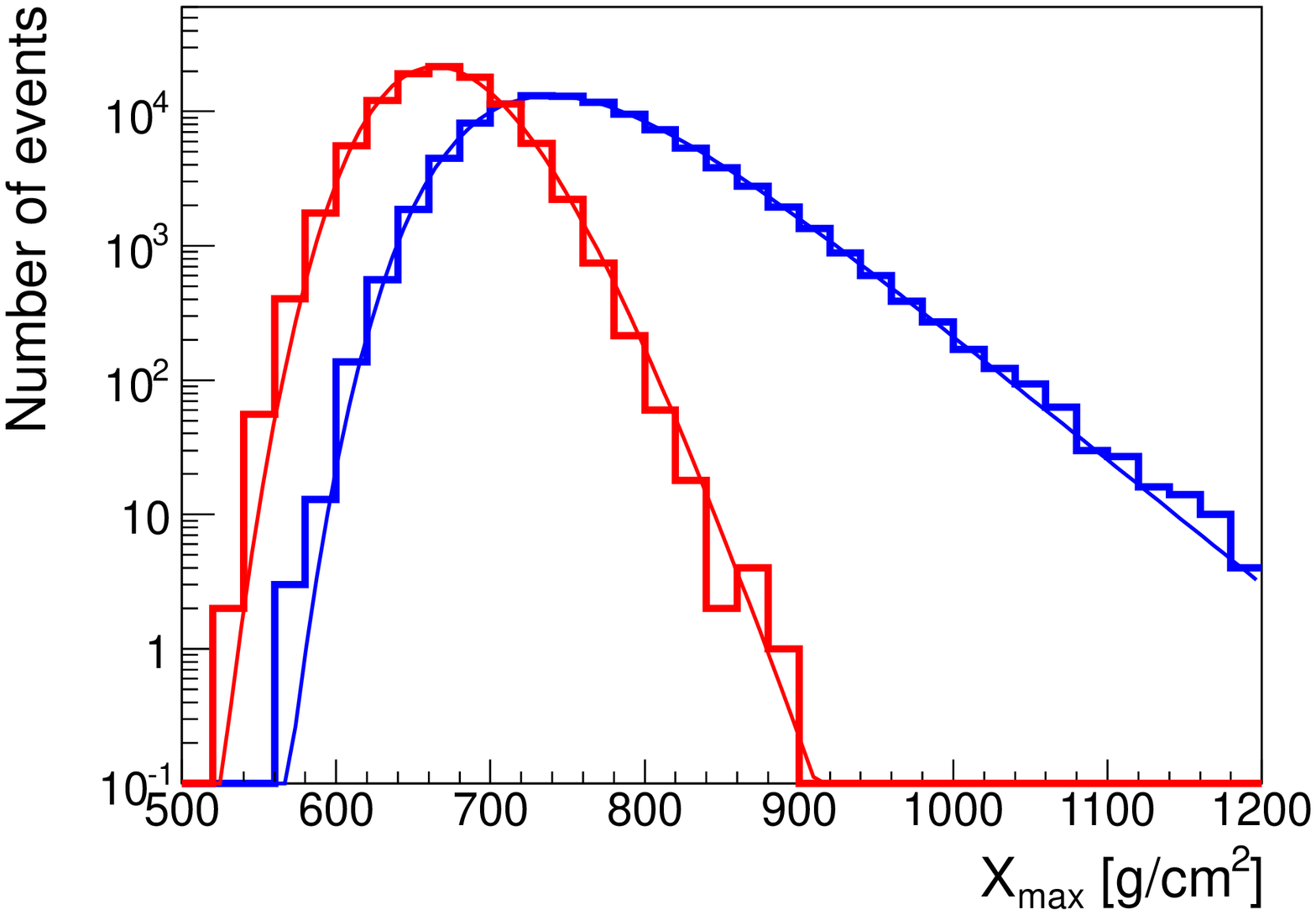}}
\subfloat{\includegraphics[width=0.5\textwidth]{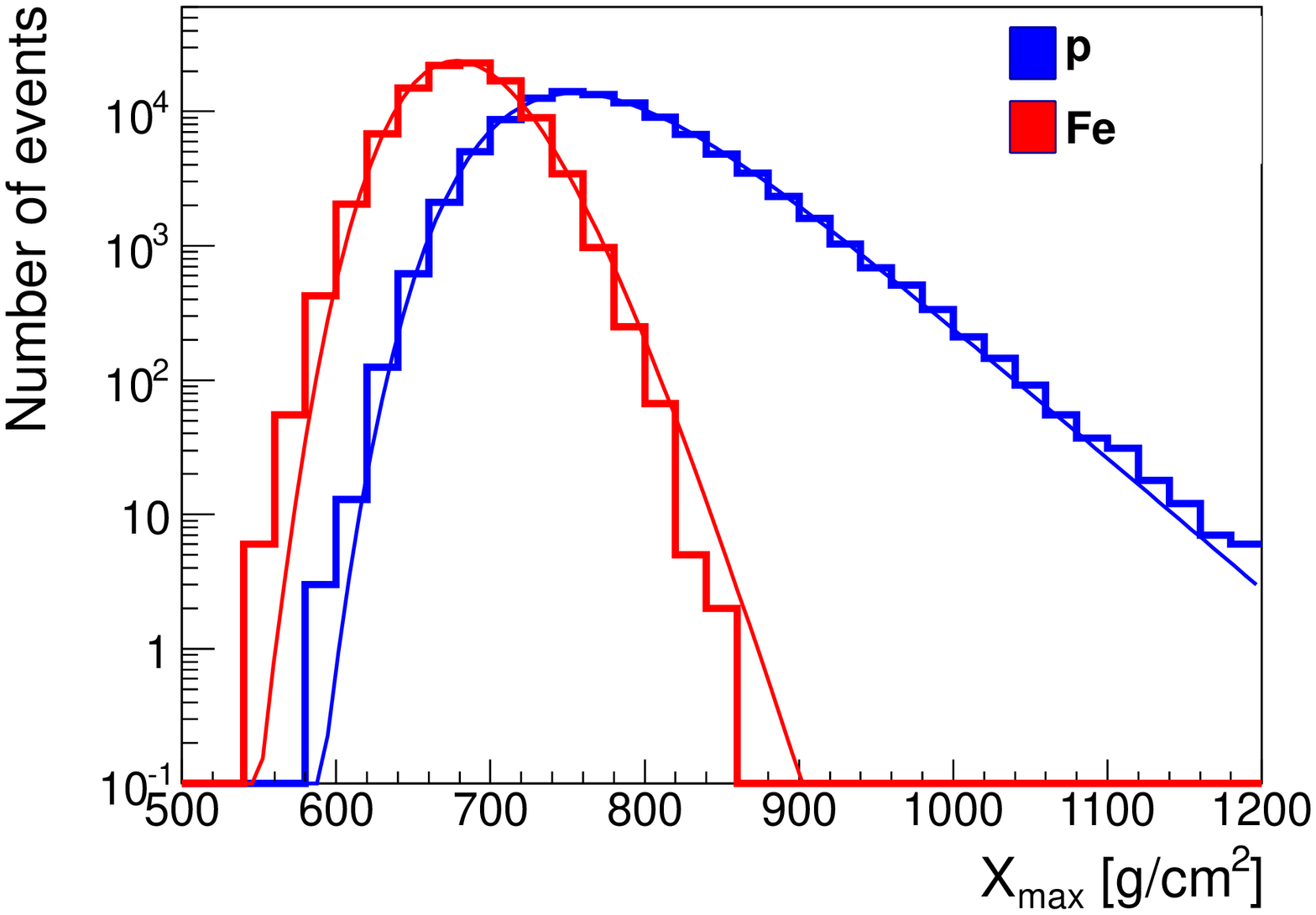}}
\caption{Distributions of $X_{\rm max}$ parametrized with Gumbel functions $g_{i}$. Showers initiated with p (blue) and Fe (red) primaries were simulated with QGSJet~II--04 (left) and EPOS--LHC (right) for $\sigma(X_{\rm max})=20$~g/cm$^{2}$.}
  \label{XmaxGumbel}
\end{figure}

\section{Method}
\label{secMethod}

\begin{figure}[ht!]
  \centering
\subfloat{\includegraphics[width=0.5\textwidth]{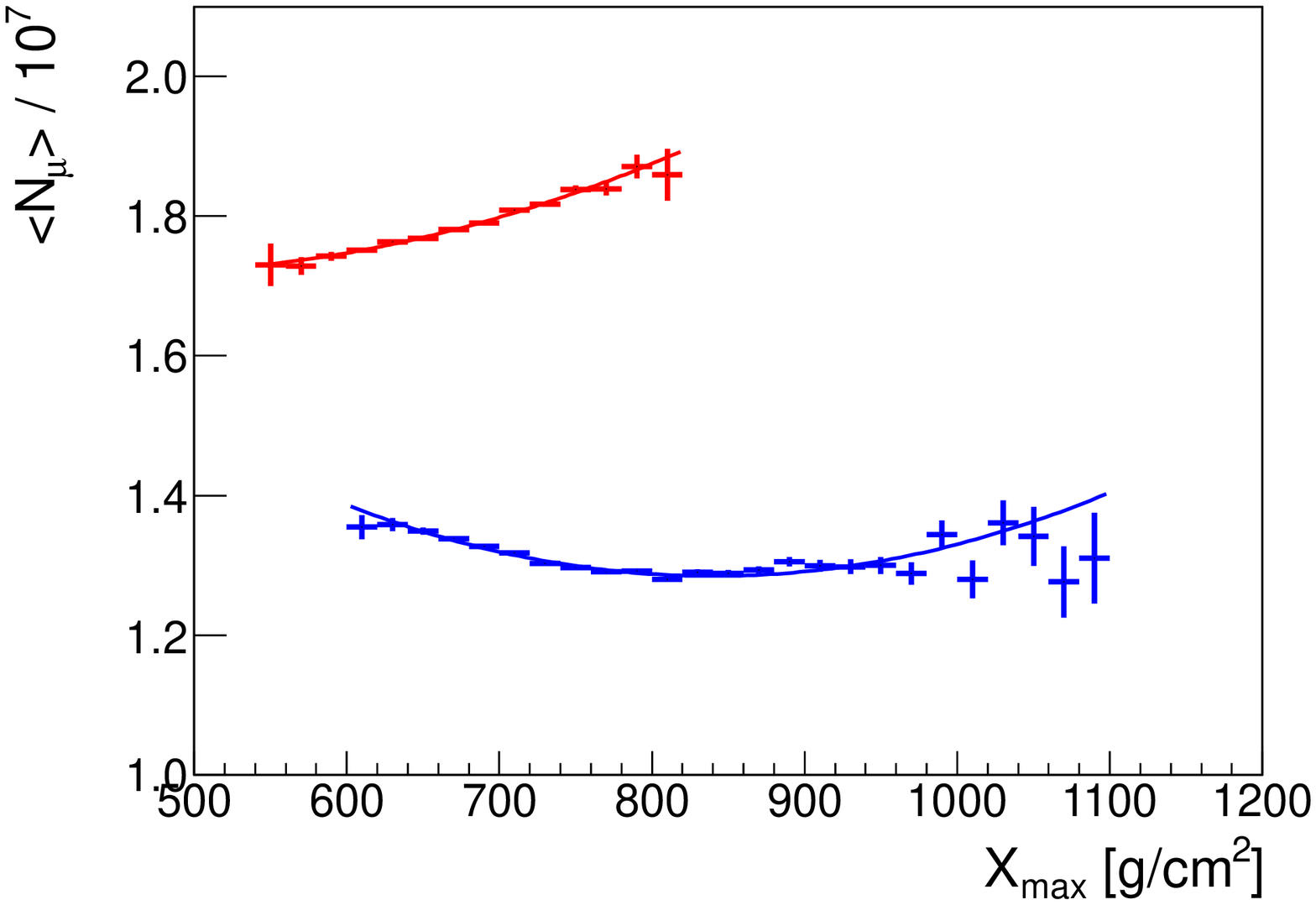}}
\subfloat{\includegraphics[width=0.5\textwidth]{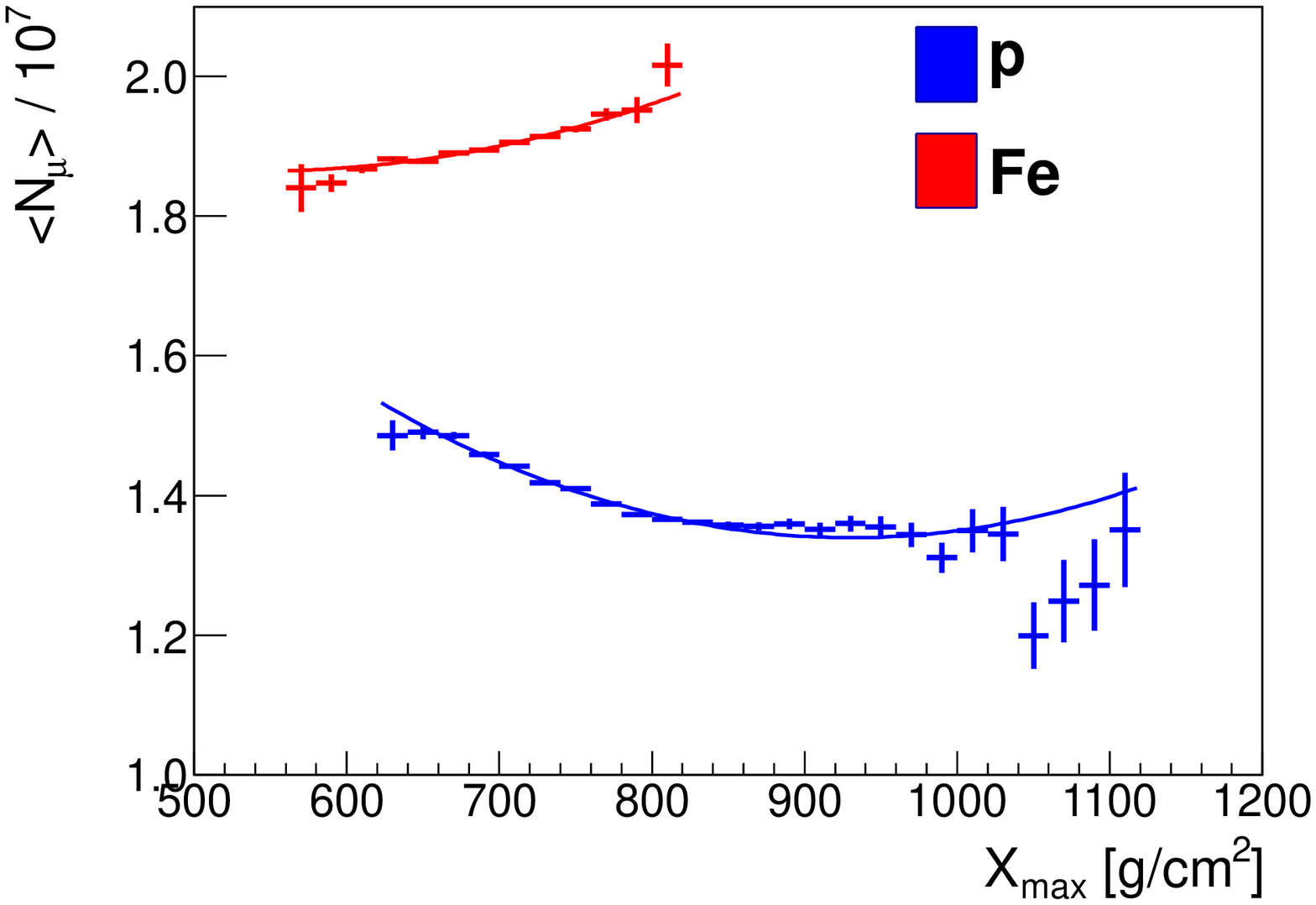}}
\caption{Dependence of mean $N_{\mu}$ on $X_{\rm max}$ parametrized with quadratic functions $<N_{\mu}^{i}>$. Showers initiated with p (blue) and Fe (red) primaries were simulated with QGSJet~II--04 (left) and EPOS--LHC (right) for $\sigma(X_{\rm max})=20$~g/cm$^{2}$ and $\sigma(N_{\mu})/N_{\mu}=10\%$. Only bins of $X_{\rm max}$ with more than 30 showers were considered in the quadratic fits.}
  \label{NmuVsXmax}
\end{figure}

For both hadronic interaction models, we parametrized $X_{\rm max}$ distributions with Gumbel functions \cite{GumbelFunction} $g_{i}$ (see Fig.~\ref{XmaxGumbel}) for both primaries $i$~=~p, Fe. We also parametrized the dependence of mean $N_{\mu}$, $<N_{\mu}>$, on $X_{\rm max}$ with quadratic functions in $X_{\rm max}$ denoted as $<N_{\mu}^{i}>$ (see Fig.~\ref{NmuVsXmax}), again for both primaries. We introduced the rescaling factor $R_{\mu}$ of $<N_{\mu}^{i}>$ to incorporate into the method the case when a rescaling of $<N_{\mu}^{i}>$ obtained from MC is needed to fit the measured $<N_{\mu}>$. Then, for a combination of two primaries with fractions $f_{i}$, $\sum f_i=1$, $<N_{\mu}>$ is given as
\begin{equation}
<N_{\mu}>=\sum_{i}\left( w_{i} <N_{\mu}^{i}>\right) R_{\mu}
\label{FitEq}
\end{equation}
where the weights $w_{i}$ are expressed as
\begin{equation}
w_{i}=\frac{f_{i}\cdot g_{i}}{\sum\limits_{j}\left( f_{j}\cdot g_{j}\right)}.
\end{equation}

For each bin of $X_{\rm max}$, $<N_{\mu}>\equiv<N_{\mu}>(X_{\rm max})$ is calculated as the weighted average of ${<N_{\mu}^{i}>\equiv<N_{\mu}^{i}>(X_{\rm max})}$ rescaling both $<N_{\mu}^{i}>$ with the same factor $R_{\mu}$. The weights ${w_{i}\equiv w_{i}(X_{\rm max})}$ reflect the relative contribution of each individual primary with relative fraction $f_{i}$ in each bin of $X_{\rm max}$ according to $g_{i}\equiv g_{i}(X_{\rm max})$.

Thus, any given dependence of $<N_{\mu}>$ on $X_{\rm max}$, which is similar to the dependence of combined proton and iron showers, can be fitted with the two--parameter ($f_{\rm p}$ and $R_{\mu}$) fit. The Fe fraction is obtained afterwards as $f_{\rm Fe}=1-f_{\rm p}$.

\begin{figure}[h!]
  \centering
\includegraphics[width=0.7\textwidth]{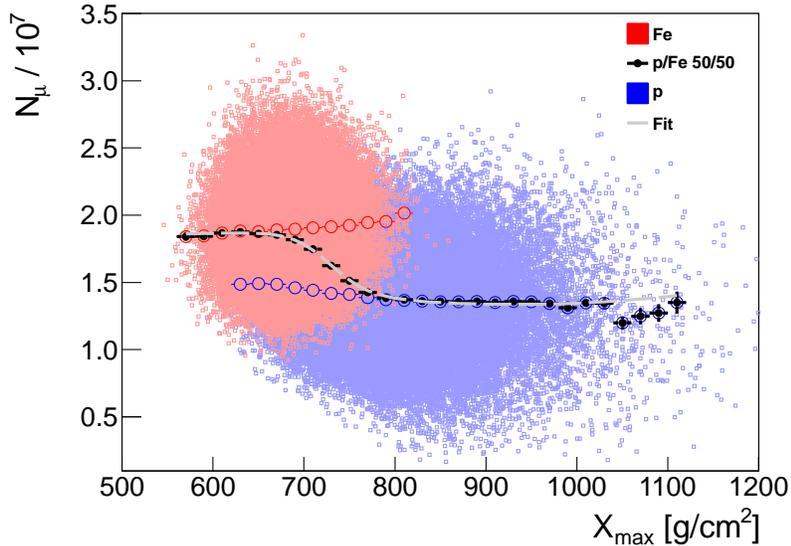}
\caption{Example of a fit (gray dashed line) for a set of showers composed of 50\% p and 50\% Fe (black points). Showers were generated with EPOS--LHC for $\sigma(X_{\rm max})=20$~g/cm$^{2}$ and $\sigma(N_{\mu})/N_{\mu}=10\%$. The individual p and Fe showers are shown with light blue and light red points, respectively. $<N_{\mu}^{\rm p}>$ is depicted with blue open points and $<N_{\mu}^{\rm Fe}>$ with red open points. The fit with $g_{i}$ and $<N_{\mu}^{i}>$ parametrized for EPOS--LHC is shown with the gray dashed line.}
  \label{FitExample}
\end{figure}

An example of application of the method to the mixed composition of showers initiated with 50\% p and 50\% Fe is shown in Fig.~\ref{FitExample}. The fitted dependence of $<N_{\mu}>$ on $X_{\rm max}$ (black points) is shown with the gray dashed line. The hadronic interaction model EPOS--LHC was considered. Starting from the lowest values of $X_{\rm max}$, $<N_{\mu}>$ matches $<N_{\mu}^{\rm Fe}>$ to about 650~g/cm$^{2}$ where a transition towards $<N_{\mu}^{\rm p}>$ begins. It continues up to about 800~g/cm$^{2}$ where $<N_{\mu}^{\rm Fe}>$ starts to match $<N_{\mu}^{\rm p}>$.

\section{Application of Method}
\label{secResults}
In this section we show basic examples of the present method and rough estimates how accurately the primary fractions of p and Fe and the muon rescaling factor can be determined. In the following, we assumed the detector resolutions to be $\sigma(X_{\rm max})=20$~g/cm$^{2}$ and $\sigma(N_{\mu})/N_{\mu}=10$\%. We considered 11 combinations of mixed compositions of p and Fe with fractions in steps of 10\% for both hadronic interaction models. Each of these compositions was reconstructed with each of the two parametrizations obtained for the two hadronic interaction models. Additionally, we reconstructed an example dataset with parametrizations of each of the two models to have another assessment of the method with respect to the two models.

\begin{figure}[h!]
  \centering
\subfloat{\includegraphics[width=0.5\textwidth]{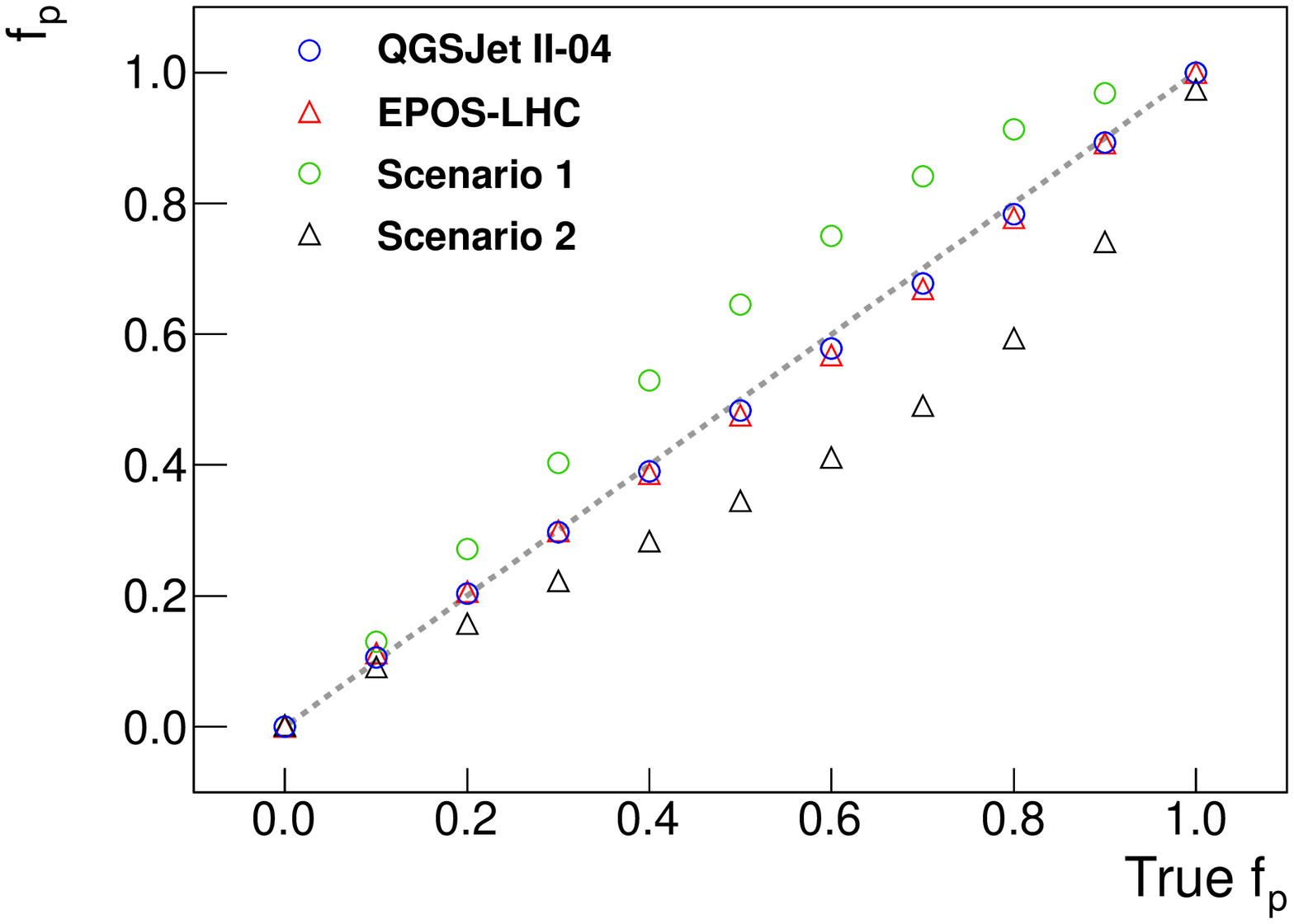}}
\subfloat{\includegraphics[width=0.5\textwidth]{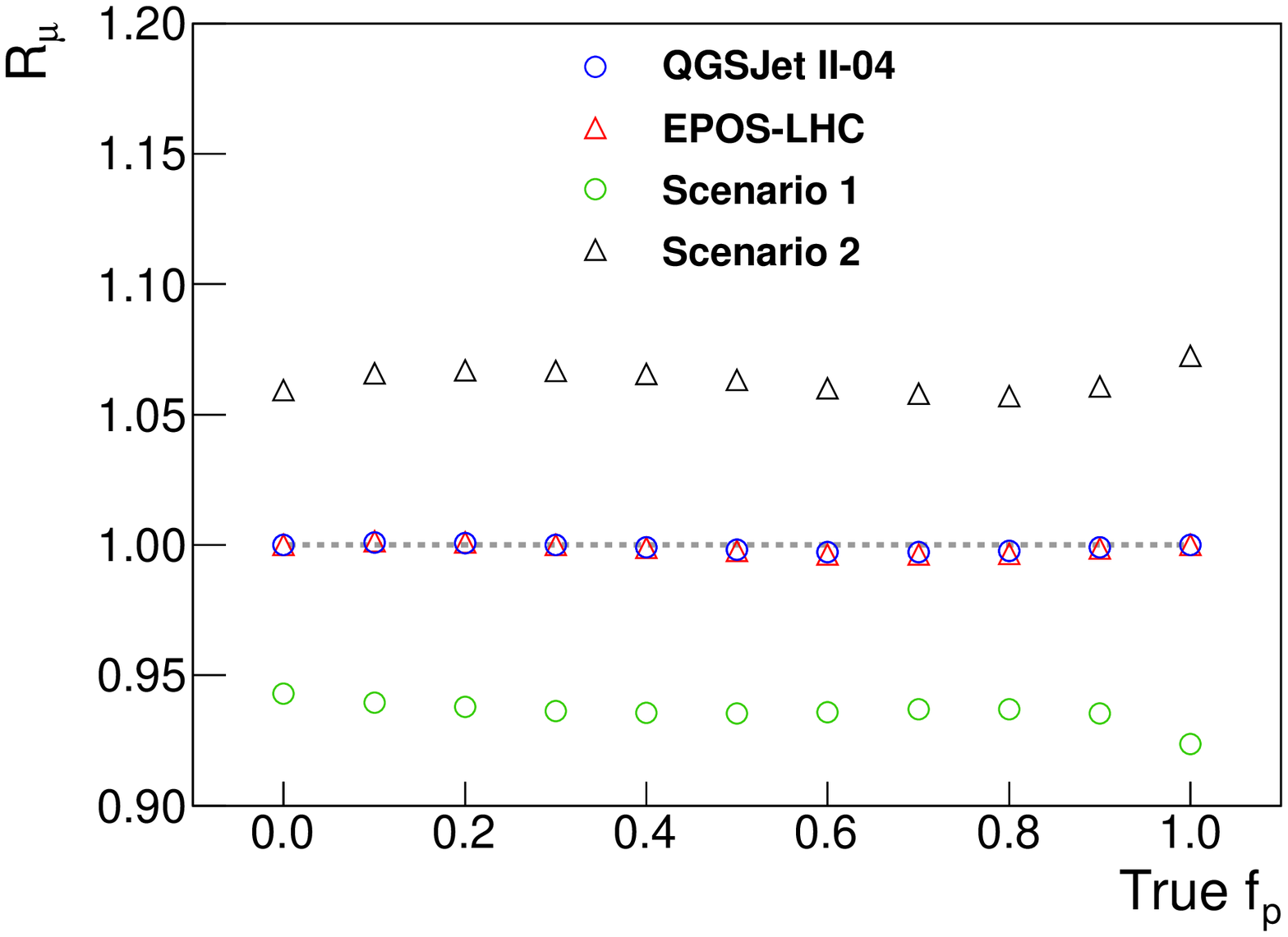}}
\caption{Left panel: Comparison of fitted ($f_{\rm p}$) and true proton fraction (True~$f_{\rm p}$). Right panel: Rescaling of $N_{\mu}$ depending on the true proton fraction. Perfect matches between fitted and true values are shown with dotted gray lines. Smearing with $\sigma(X_{\rm max})=20$~g/cm$^{2}$ and $\sigma(N_{\mu})/N_{\mu}=10\%$ was considered. Scenario 1 (2) corresponds to QGSJet~II--04 (EPOS--LHC) showers fitted with parametrizations of $g_{i}$ and $<N_{\mu}^{i}>$ from EPOS--LHC (QGSJet~II--04).}
  \label{Comparisons}
\end{figure}

On the left panel of Fig.~\ref{Comparisons}, a comparison of the fitted and the true proton fraction is shown for QGSJet~II--04 (blue) and EPOS--LHC (red). The reconstructed $f_{\rm p}$ of showers generated with a different model than that used for the parametrization of $g_{i}$ and $<N_{\mu}^{i}>$ are depicted by open black markers. Scenario 1 (2) corresponds to showers produced with QGSJet~II--04 (EPOS--LHC) and fitted with the parametrizations obtained from EPOS--LHC (QGSJet~II--04) showers. 

When the same hadronic interaction model is used for the generation of showers and parametrization of $g_{i}$ and $<N_{\mu}^{i}>$, the proton fraction is reconstructed within a few \% of the true value. However, in cases of Scenario 1 and 2, the difference between the fitted and the true proton fraction increases with the spread of primary masses of the selected composition. It reaches values up to $\sim$20\%.

On the right panel of Fig.~\ref{Comparisons}, the muon rescaling is plotted for different true proton fractions. The rescaling factor is found to be within a few \% to 1 (precision of the method), when the same models were used for parametrization and generation (red and blue). Black points correspond to the relative difference of $N_{\mu}$ for showers generated with QGSJet~II--04 and EPOS--LHC, which is about 6\%.

\begin{figure}[h!]
  \centering
\includegraphics[width=0.7\textwidth]{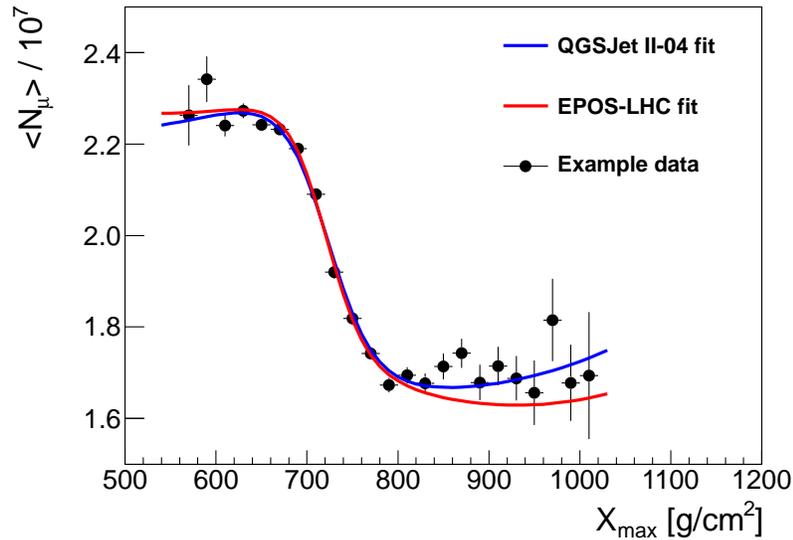}
\caption{An example of data (black points) fitted with QGSJet~II--04 (blue) and EPOS--LHC (red) parametrizations.}
  \label{FittedExample}
\end{figure}

As another check, we created example data from 5000 p and 5000 Fe showers produced with QGSJet~II--04. For each shower, we scaled $N_{\mu}$ by a factor 1.3 and increased $X_{\rm max}$ by 7~g/cm$^{2}$. Note that EPOS--LHC generates showers with deeper $X_{\rm max}$ than QGSJet~II--04 by about 14~g/cm$^{2}$ on average. These example data were fitted with parametrizations of both models (see Fig.~\ref{FittedExample}). Both fits describe the example data similarly well giving different proton fractions and muon rescaling factors that are shown in Tab.~\ref{FitsTable}. The difference of $f_{\rm p}$ is about 15\% when different parameterizations (Figs.~\ref{XmaxGumbel},\ref{NmuVsXmax}) based on two most recent models are used. The ratio of $R_{\mu}$ for the two models reflects again that EPOS--LHC produces about 6\% more muons than QGSJet~II--04 on average.

\begin{table}[!h]
\begin{center}
\caption{Fitted parameters for example data.}
\begin{tabular}{|c|c|c|}
\hline
Model & $f_{\rm p}$ [\%] & $R_{\mu}$\\
\hline\hline
QGSJet~II--04 & 41 $\pm$ 2 & 1.297 $\pm$ 0.004 \\
\hline
EPOS--LHC & 56 $\pm$ 2 & 1.216 $\pm$ 0.004 \\
\hline

\end{tabular}
\label{FitsTable}
\end{center}
\end{table}

\section{Conclusions}
A method of simultaneously obtaining the primary fractions and the muon rescaling factor from $X_{\rm max}$ and $N_{\mu}$ of UHECR was presented. Simulated showers with two models of hadronic interactions tuned to LHC data (run I-II) were used. The precision of the method was tested with different combinations of p and Fe primaries and with example data. The primary fractions and the muon rescaling factor can be determined within a few \%. The difference of the proton fraction reconstructed with the two parameterizations based on the two models of hadronic interactions was observed below 20\%. The muon rescaling factor reflected the relative difference (around 6\%) in the average muon shower size of the two models of hadronic interactions.

\section*{Acknowledgements}

This work is funded by the Czech Science Foundation grant 14-17501S.

\end{document}